\documentclass[12pt]{article}
\usepackage{mathptmx,graphicx,amsmath,amssymb,amsfonts,latexsym, dsfont}
\usepackage{subfigure, units}
\usepackage{color}
\usepackage{cancel}

\begin{document}
\begin{center}
{\Large\bfseries
One- and two-dimensional photonic crystal micro-cavities in single crystal diamond \par}
\vspace{3ex}
{\bfseries
Janine Riedrich-M\"{o}ller$^1$, Laura Kipfstuhl$^1$, Christian Hepp$^1$, Elke Neu$^1$, Christoph Pauly$^2$,    Frank M\"ucklich$^2$,
Armin Baur$^3$, Michael Wandt$^3$, Sandra Wolff$^4$,
Martin Fischer$^5$, Stefan Gsell$^5$, Matthias Schreck$^5$,  and
Christoph Becher$^{1}$\footnote{Corresponding author: {\emph{christoph.becher@physik.uni-saarland.de}},  Tel.: +49 (0)681 302 2466; Fax: +49 (0)681 302 4676}
}
\end{center}
{\footnotesize\itshape
1. Universit\"{a}t des Saarlandes, Fachrichtung 7.2 (Experimentalphysik), 66123 Saarbr\"{u}cken, Germany \\
2. Universit\"{a}t des Saarlandes, Fachrichtung 8.4 (Materialwissenschaft und Werkstofftechnik), 66123 Saarbr\"{u}cken, Germany \\
3. University of Freiburg, Departement of Microsystems Engineering (IMTEK), Cleanroom Service Center, 79110 Freiburg, Germany \\
4. TU Kaiserslautern, Nano+Bio Center, 67653 Kaiserslautern, Germany \\
5. Universit\"{a}t Augsburg, Lehrstuhl f\"ur Experimentalphysik IV, 86159 Augsburg, Germany}
\vspace{3ex}

\textbf{The development of solid-state photonic quantum technologies is of great interest for fundamental studies of light-matter interactions and quantum information science. Diamond has turned out to be an attractive material for integrated quantum information processing due to the extraordinary properties of its colour centres enabling e.g. bright single photon emission and spin quantum bits. To control emitted photons and to interconnect distant quantum bits, micro-cavities directly fabricated in the diamond material are desired. However, the production of photonic devices in high-quality diamond has been a challenge so far. Here we present a method to fabricate one- and two-dimensional photonic crystal micro-cavities in single-crystal diamond, yielding quality factors up to 700. Using a post-processing etching technique, we tune the cavity modes into resonance with the zero phonon line of an ensemble of silicon-vacancy centres and measure an intensity enhancement by a factor of 2.8. The controlled coupling to small mode volume photonic crystal cavities paves the way to larger scale photonic quantum devices based on single-crystal diamond. }

A number of seminal experiments have demonstrated the prospects of colour centres in diamond, in particular the negatively charged nitrogen-vacancy centre (NV$^{-}$), for application as spin quantum bit (qubit) in quantum information processing \cite{Ladd2010}. Local coupling between NV$^{-}$ qubits can be mediated by magnetic interactions \cite{Neumann2010}, whereas long-distance coupling might rather be promoted by optical interactions \cite{Benjamin2009}. Such transfer of quantum information  between spins and photons in solid state systems \cite{Togan2010} can be strongly enhanced by optical micro-cavities with small mode volume $V_{\text{mod}}$ and large quality factor $Q$. Micro-cavities can further be employed for enhancing the emission rate of single photon emitters \cite{Noda2007, Su2008} based e.g. on NV$^{-}$ centres \cite{Kurtsiefer2000} or bright Cr-related complexes \cite{Aharonovich2009b} and silicon-vacancy (SiV) centres \cite{Neu2011}, for cavity enhanced spin measurements \cite{Young2009} and quantum communication \cite{Childress2006}. For all these schemes it is important to selectively enhance the zero phonon line (ZPL) emission and suppress emission into phonon side-bands and non-radiative decay channels. Micro-cavities based on photonic crystal (PhC) structures provide both very small mode volumes for strong emitter-cavity coupling and scalable architectures for integrated photonic networks \cite{Greentree2008}. In recent years, the design of high-Q diamond based PhC cavities has been studied theoretically, yielding Q-factors in the range of  10$^4$-10$^6$ for two-dimensional point defect cavities \cite{Tomljenovic-Hanic2006a, Kreuzer2008, Riedrich-Moller2010} and heterostructure cavities \cite{Bayn2006, Tomljenovic-Hanic2009} as well as one-dimensional waveguide cavities \cite{Babinec2011}.

For experimental realisation of colour centre-cavity coupling two main approaches have been pursued. The first route is to assemble hybrid systems where colour centres in diamond nanocrystals or bulk diamond are coupled to the evanescent fields of cavities defined in non-diamond materials for which established nano-fabrication techniques exist: here coupling to silica micro-spheres \cite{Park2006, Schietinger2008}, silica micro-disks \cite{Barclay2009}, GaP micro-disks \cite{Barclay2009b} and GaP micro-ring cavities \cite{Fu2011} has been demonstrated. In a very similar fashion recent experiments have realised controlled coupling of NV$^{-}$ centres in nanodiamonds to GaP photonic crystal cavities \cite{Wolters2010, Englund2010a, Sar2011}. In these experiments selective enhancement of the NV$^{-}$ ZPL could be detected in the cavity emission spectrum. The hybrid coupling scheme, however, is constrained by several limitations, e.g. spatial mismatch between emitter and cavity field maximum, scattering and losses from the cavity materials as well as line-broadening mechanisms such as spectral diffusion or random strain broadening for single NV$^{-}$ centres in diamond nanocrystals  \cite{Shen2008, Santori2010} or  the lack of NV$^{-}$ fluorescence close to the  surface of bulk diamond \cite{Santori2010}.

Given these constraints, single crystal diamond is the material of choice as colour centres with best possible  properties, i.e. narrow optical emission linewidth \cite{Tamarat2006} and long  spin coherence lifetimes \cite{Balasubramanian2009},  can be directly integrated into photonic structures. As there is no need for evanescent coupling colour centres can be placed at sufficient distance from the surface to minimise influence of surface impurities or defects. This second approach to colour centre-cavity coupling, however, has proven to be very challenging as techniques for fabrication of thin high-quality diamond films and for diamond patterning at the nanoscale are not readily available.
In first experiments, micro-disks \cite{Wang2007a} and PhC cavities \cite{Wang2007} have been fabricated  in nano-crystalline diamond films, grown on a sacrificial substrate. Compared to single crystal diamond, however, the optical quality of  nano-crystalline diamond films is inferior due to intrinsic scattering and absorption losses. A technique for the production of free-standing, mono-crystalline diamond membranes is graphitisation by high energy ion-implantation producing a sacrificial layer which allows for lifting out a thin membrane \cite{Fairchild2008}. Yet, residual damage due to the ion-implantation causes high optical losses and significant modification of the colour centre (NV$^{-}$) luminescence \cite{Bayn2011a}. PhC cavities have been fabricated in these membranes using focused ion beam (FIB) milling \cite{Bayn2011} but due to the inferior material quality no cavity modes could be observed. Very recently, one-dimensional PhC cavities \cite{Bayn2011a, Babinec2011}  and micro-ring resonators \cite{Faraon2010}  have been fabricated in high-quality single crystal diamond by FIB milling and dry etching, respectively obtaining Q-factors of $\sim$ 220 \cite{Bayn2011a} and 4300  \cite{Faraon2010}. For the micro-ring cavity, controlled coupling to individual NV$^{-}$ centres was achieved, resulting in a spontaneous emission lifetime reduction and spectral enhancement of the ZPL.

Here, we present a novel method to realise one- and two-dimensional PhC cavities in single crystal diamond films grown on a sacrificial substrate. To this end, we use a unique material system \cite{Gsell2004} which enables  low-damage fabrication of thin, free-standing membranes. Furthermore, we develop an extensive post-process cleaning procedure to remove the damage induced by FIB nanopatterning. The combination of the two approaches allows us to observe for the first time high-Q modes in PhC cavities defined in single crystal diamond.
As starting material we use heteroepitaxial diamond films with a thickness of about \unit[12]{$\mu$m} grown by chemical vapour deposition on silicon(001) substrates via iridium/yttria-stabilised zirconia buffer layers \cite{Gsell2004}.
In a first step, we fabricate a free-standing  membrane by removing the silicon substrate and buffer layers in small areas of the sample.
In a second step, the  diamond film is  thinned down from the nucleation side to a thickness of \unit[300]{nm} using reactive ion etching (RIE) in an oxygen plasma.
After the membrane preparation, the PhC structure is milled into the diamond film by a focused beam of Ga$^{+}$-ions (FIB).
To recover a high quality diamond layer, the sample is annealed and cleaned after the milling process. Fabrication details are provided in the Methods section.

\begin{figure}[h!]
  \centering
  \includegraphics[width=1\textwidth]{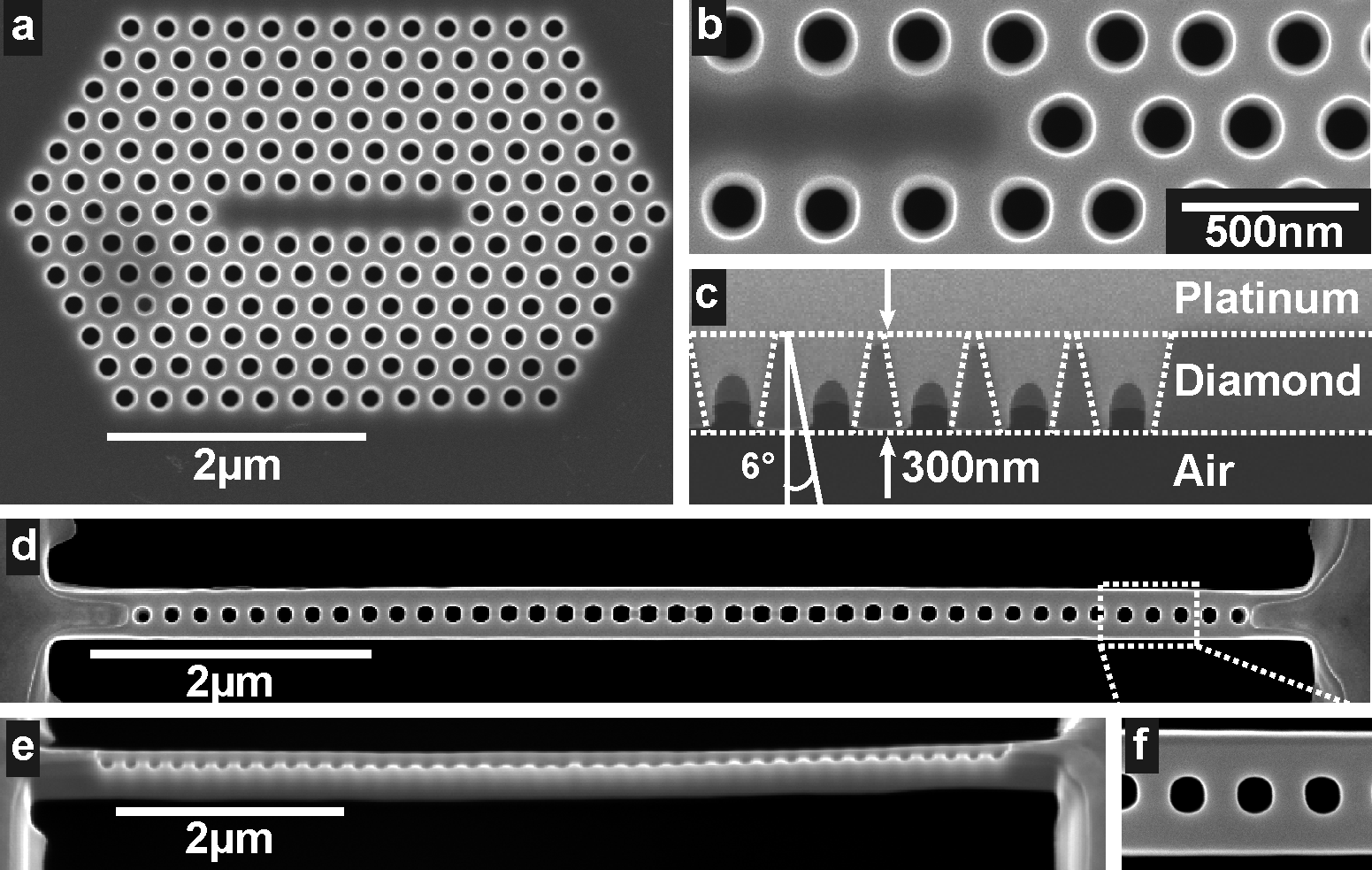}
  \caption{Scanning electron microscope (SEM) images of 2D and 1D fabricated PhC cavities: (a) SEM image of the fabricated M7-cavity  with lattice constant $a \approx $\unit[275]{nm} and radii $R \approx $\unit[85]{nm}. (b) Close-up view of the cavity centre. (c) Cross-sectional image (tilt angle: 52$^{\circ}$): a thin platinum layer was deposited on the photonic crystal in order to allow for a straight cut through the diamond membrane using FIB. From the cross-sectional image, a diamond film thickness of \unit[300]{nm} can be inferred. The sidewalls of the milled air holes exhibit a tilt angle of about 6$^{\circ}$. (d) Top-view and (e) side-view  of the fabricated 1D nanobeam cavity with  a pitch:width:height ratio of 2:3:3 ($a \approx $ \unit[200]{nm}). The hole radii monotonically decrease from $R \approx $ \unit[83]{nm} at the cavity centre to $R \approx $ \unit[72]{nm} at the waveguide edge.  (f) Close-up view of the 1D waveguide-cavity.
  \label{fig:REM}}
\end{figure}

We fabricate both one- and two-\-dimensional PhC cavities (Fig. \ref{fig:REM}): The photonic structures are designed such that the high-Q cavity modes lie close to the emission wavelengths of well-investigated colour centres, i.e. NV$^{-}$ and SiV centres with applications as quantum bit and as bright single photon emitters, respectively. SiV centres are incorporated in the diamond film during the  growth process \cite{Neu2011}, whereas  NV$^{-}$ centres are not observed in the sample. However, there exist techniques to deterministically implant NV centres after the patterning process at well defined positions in the PhC cavities \cite{Meijer2008}.
The two-dimensional photonic crystals  consist of a triangular lattice of air holes with a seven-hole-defect  (M7-cavity). These simple defect structures were chosen as ``proof-of-principle'' systems as they provide several cavity modes facilitating comparison between theory and experiment. Theoretical quality factors are up to 11,800  for the fundamental mode with a mode volume of $V_{\text{mod}} = $ \unit[1.5]{$(\lambda/n)^3$}, where $n = 2.4$ is the refractive index of diamond.
As a more elaborated design, we fabricate  one-\-dimensional integrated waveguide-cavity structures (nanobeam cavities).
They consist of a suspended ridge waveguide with a line of uniformly spaced air holes. A cavity is introduced in the centre of the nanobeam by deterministically varying the size of the holes according to the  ``modulated Bragg'' design \cite{Quan2010}: The filling factor $FF$ of each nanobeam segment linearly decreases from the cavity centre to the edge of the structure. Here the filling factor is defined as the ratio of the hole area to the area of each unit cell. This deterministic design tailors the cavity field in such a way that  it gently decays along the waveguide-cavity. The gentle confinement \cite{Yoshihiro2003} minimises radiation losses,  yielding theoretical quality factors up to $6\cdot 10^5$ with a very small mode volume $V_{\text{mod}} = 0.7 (\lambda/n)^3$.

Figure \ref{fig:REM} shows  scanning electron microscope (SEM) images of the fabricated PhC structures. The lattice constant $a \approx$ \unit[275]{nm} and air hole radius $R \approx$ \unit[85]{nm} of the two-dimensional M7-cavity (Fig. \ref{fig:REM}(a,b)) are chosen such that the cavity modes are close to resonance with the ZPL of the SiV centres ($\lambda = $\unit[738]{nm}). The cross-sectional SEM image (Fig. \ref{fig:REM}(c)) reveals slightly tilted air hole  sidewalls with a tilt angle of about $6^{\circ}$
possibly resulting from the Gaussian cross section of the focused ion beam. From the cross-sectional image we determine a diamond film thickness of $h \approx$ \unit[300]{nm}.
The nanobeam cavity with a lattice constant  $a \approx $ \unit[200]{nm} is designed to form modes resonant with the ZPL of NV$^{-}$ centres ($\lambda = $ \unit[637]{nm}). Figure \ref{fig:REM}(d-f)  show the top- and side-view SEM images  of the fabricated one-dimensional waveguide-cavity structure with a thickness $h$ equal to the width $w$ of \unit[300]{nm}. The hole radii decrease from $R \approx $ \unit[83]{nm} at the cavity centre to $R\approx$ \unit[72]{nm} at the edge of the structure.

We use a confocal microscope setup with a numerical aperture of 0.8  for optical excitation and collection of  the diamond fluorescence. A continuous wave (cw) laser at \unit[532]{nm} with \unit[600]{$\mu$W}  is used to excite the structures. The room-temperature photoluminescence (PL) spectrum  of a M7 PhC cavity  and a reference spectrum of the unstructured membrane  are shown in Figure \ref{fig:Modenbilder}.
In both spectra we observe  a pronounced ZPL of SiV centres embedded within the cavity at a wavelength of $\lambda = $ \unit[738]{nm}. At longer wavelengths, in the range of   $\lambda = $ \unit[745-825]{nm}, multiple cavity modes are observed in the M7-cavity spectrum, which are not present in the reference spectrum of the unstructured membrane.
Figure \ref{fig:Nanobeam-spec} shows the PL spectrum of a nanobeam cavity and a reference spectrum of the bare membrane. Three distinct resonant modes in the range of $\lambda = $ \unit[613-640]{nm} are observed in the cavity spectrum.
Note that the cavity modes above ($\lambda \gtrsim  $ \unit[750]{nm}) and below ($\lambda \lesssim  $ \unit[730]{nm}) the SiV ZPL are fed by broadband luminescence of the diamond material.

In order to identify the cavity modes, we perform finite-difference time-domain (FDTD) simulations. To this end, we apply different symmetric boundary conditions to calculate the expected frequencies of the fundamental mode $e1$ and higher order modes  for an ideal M7-cavity (Fig. \ref{fig:Modenbilder}) and a nanobeam cavity (Fig. \ref{fig:Nanobeam-spec}). The calculated modes are in good agreement with the experimental spectra.
The small discrepancy between the experimental and theoretical spectra  results from fabrication tolerances (see supplementary information).

The simulations also allow to analyse the polarisation of the cavity modes.  The $E_x$ and $E_y$ mode profiles of the fundamental mode $e1$ and higher order modes $e2$, $o1$ and $o2$ are shown in Figure \ref{fig:Spektren}(a,b) for the M7-cavity and nanobeam cavity, respectively. According to the symmetry of the mode profiles with respect to the mirror plane at $y = 0$ (centre line of the cavity), the modes can be classified in even ($e$) and odd ($o$) modes.
In the far field the component of the electric field cancels, which has a node along the cavity centre line \cite{Kim2004}.
For even (odd) modes the $E_x$- ($E_y$-) component vanishes in the far field and we would expect  the even (odd) modes to be linearly polarised in the $y$- ($x$-) direction.
The polarisation measurements of the fabricated PhC cavities are shown in Figure \ref{fig:Spektren}(c,d): The black curve is measured without a polarisation analyser whereas the red and blue curves are spectra taken with an analyser oriented in the $y$- or $x$-direction, respectively.
The even modes of the M7-cavity (Fig. \ref{fig:Polarisation}) are pronounced for a  polarisation analyser along the $y$-axis, whereas the odd modes are  visible  for an analyser in the $x$-direction.
In the case of the integrated waveguide-cavity structure, the polarisation analysis (Fig. \ref{fig:Nanobeam-pol}) reveals that all cavity modes are linearly polarised in the $y$-direction. This  is expected  as the  observed nanobeam cavity modes originate from the same dielectric-band of an unperturbed waveguide mode showing an even symmetry.
The polarisation measurements are in good agreement with the theoretical predictions (see also supplementary information) and  unambiguously prove that the observed spectral features are true  cavity modes.

From the measured spectrum we obtain the quality factors $Q = \lambda_c/\Delta \lambda_c$, where $\lambda_c$ is the resonance wavelength and $\Delta \lambda_c$ is the corresponding peak width (FWHM).
The experimental and theoretical Q-factors of the lowest energy modes are listed in Tables \ref{tab:Q-M7} and \ref{tab:Q-Nanobeam}. The highest experimental Q-factors  $Q_{\text{Exp}} $ are 700 and 450 for the nanobeam and M7-cavity, respectively.
Compared to the theoretical quality factors $Q_{\text{Theory}}$ of the fundamental modes, the measured quality factors $Q_{\text{Exp}}$ are one to three orders of magnitude smaller. There is just on exemption: the experimental Q-factor of the $o2$-mode  is even larger compared to theoretical predictions (see Tab.\ref{tab:Q-M7}).
The strong degradation of the Q-factors is mainly due to the non-vertical sidewalls of the air holes \cite{Babinec2011, Bayn2011}. Table \ref{tab:Q-M7} lists the calculated Q-factors $Q_{\text{Tilt}}$ of an otherwise ideal M7-cavity but with inclined  sidewalls (tilt angle: 6$^{\circ}$). The cone shape of the air holes leads to a reduction of the maximal Q-factor to 900, which primarily affects the high-Q $e1$ and $e2$ modes. The Q-factors are further influenced by the incidental departure of the fabricated cavity structures from the ideal in-plane geometry. From SEM images we infer an average deviation of air hole positions and radii from their ideal values of $\sim$ \unit[13]{nm} and $\sim$ \unit[6]{nm}, respectively. To further quantify the impact of fabrication tolerances, we convert the SEM picture of the fabricated M7-cavity into a dielectric structure and calculate the Q-factors using FDTD simulations (see supplementary information). Table \ref{tab:Q-M7} lists the quality factors $Q_{\text{SEM}}$ of the M7-cavity with a geometry based on the imported SEM image (note that due to software limitations we have to assume vertical sidewalls here). The quality factor of the fundamental mode $e1$ drops by a factor of $\sim 3$ whereas the Q-factors of the $e2$ and $e3$ modes are almost unaffected. Surprisingly, the $Q_{\text{SEM}}$ factors of the odd modes are higher than for the ideal M7-cavity. This is due to an incidental variation of the air holes surrounding the cavity centre having a strong impact on the odd cavity modes.
Absorption losses of the single crystal diamond material play only a minor  role, as the absorption coefficient measured for a comparable material system is $<$\unit[10]{cm$^{-1}$} (see supplementary information).

In both fabricated cavities the resonant modes are spectrally very close to the ZPLs of diamond colour centres (NV$^{-}$ in case of nanobeam cavity, SiV in case of M7-cavity).
In general, however, due to fabrication tolerances it is challenging to fabricate cavity structures whose resonance wavelength matches exactly the emission line of colour centres in the diamond film. Therefore, it is indispensable to actively tune the cavity after the fabrication process. To this end we make use of the fact that the resonance wavelengths of the PhC cavities are very sensitive to structural parameters like the slab thickness $h$ and hole radius $R$. Simulations of the PhC  cavity structure reveal that the modes shift to shorter wavelengths for both smaller thickness $h$ and larger radii $R$.
Notably the radius has a strong impact on the cavity  wavelengths: Increasing the radius by only \unit[1.5]{nm}  results in a blue-shift of the cavity resonances by $\sim$\unit[3]{nm} (see supplementary information). This observation motivates a ``digital etching'' tuning technique, very much in the spirit of the method devised for semiconductor PhCs \cite{Hennessy2005}.
We demonstrate this technique for the fabricated two-dimensional cavity: In the spectrum shown in Figure \ref{fig:Modenbilder}, the M7-cavity modes are red-shifted with respect to the SiV ZPL.
In order to blue-shift the cavity modes, we repeatedly  oxidise the sample for \unit[10]{min} at 480$^{\circ} $C  \cite{Osswald2006} in air to slightly decrease the slab thickness and increase the size of the air holes.
This modification of the structure parameters leads to a blue-shift of the M7-cavity modes by $\sim$\unit[3]{nm} on average per oxidation step (Fig. \ref{fig:D115-CavityTuning}),
which is in good agreement with the theoretical predictions (see supplementary information).  In total, the cavity modes are shifted up to  \unit[15]{nm} (Fig. \ref{fig:l-Oxid})  without  significant degradation of  the  Q-factors (Fig. \ref{fig:Q-Oxid}). In a similar manner the nanobeam resonant modes are blue-tuned by $\sim$\unit[3]{nm} on average per oxidation step and  up to a total of \unit[14]{nm} after five oxidation steps.
By oxidising the diamond film, we are able to spectrally tune the M7-cavity mode $o2$ in resonance with the ZPL of an ensemble of SiV centres (Fig. \ref{fig:D115-CavityTuning}).  On resonance, we observe a clear enhancement of the intensity of  the SiV ZPL by a factor of 2.8 compared to the off-resonant spectrum.
This increase of the intensity is due to a spectrally resolved Purcell enhancement of the SiV centres' spontaneous emission. Considering the quality factor $Q=400$ and the mode volume $V_{\text{mod}} = $ \unit[1.5]{ $(\lambda/n)^3$} of the $o2$ mode, an ideal Purcell factor $ F_c = 3/4\pi^2 (\lambda/n)^3 Q/V_{\text{mod}} = 20$ can be deduced as the cavity figure of merit. A more detailed analysis (see supplementary information) has to take into account spatial and oriental averaging of the emitter-mode overlap; spectral mismatch of SiV ZPL width and cavity linewidth;  modification of the local density of states by the photonic crystal and different collection efficiencies for cavity mode and un-coupled emission.  Furthermore, one has to account for the branching ratio of SiV emission into ZPL, phonon sidebands and non-radiative decay channels, which is approximately  4\%:1\%:95\% \cite{Neu2011}.
This analysis predicts an  enhancement of the ZPL intensity of $3\pm1$, which is in good agreement with the experimental measurements. As this spontaneous emission Purcell enhancement is averaged over the spatial and spectral distribution of the SiV centre ensemble, the total spontaneous emission rate is virtually unchanged by the cavity coupling (see supplementary information). Note, however, that for an emitter with perfect spatial and spectral matching to the cavity mode, a Purcell factor of $F_c = 20$  results in an approximate two-fold emission rate enhancement and a fraction of $\approx 50\%$ of total emission into the ZPL, thereby increasing the radiative quantum yield by a factor of 10. Such an experiment would require a deterministic placement of a single SiV centre relative to the cavity mode which is within reach of current technologies.

In summary, we have  fabricated for the first time one- and two-dimensional PhC cavities in single crystal diamond films yielding quality factors up to  700 with  mode volumes between $0.7 (\lambda/n)^3$ and $1.5 (\lambda/n)^3$. The observed cavity modes are spectrally close to the ZPLs of prominent colour centres (NV$^{-}$, SiV) established as quantum bits or single photon sources. By post-processing the fabricated structures,  we are able to spectrally shift the resonant modes up to \unit[15]{nm} and tune a two-dimensional cavity  into resonance with the emission line of SiV colour centres embedded in the diamond film. On resonance, we observe a clear increase in the ZPL intensity due to a spectrally resolved Purcell enhancement. The very same tuning procedure can be used in principle to tune the nanobeam cavity modes into resonance with the  ZPL of NV$^{-}$ centres that could be implanted at well defined positions into the cavity structure subsequently to  patterning.
The currently achieved Q-factors  are found to be limited by the precision of the PhC fabrication technique rather than by the optical quality of the heteroepitaxial diamond films.
As a consequence, processing techniques to pattern the diamond membrane such as electron beam lithography and reactive ion etching which allow for smaller geometric tolerances and reduced material damage than the focused ion beam method should enable fabrication of PhCs with significantly higher Q-factors in this technologically interesting diamond material.
These techniques might enable realisation of large scale photonic designs and integrated  cavity-waveguide structures \cite{Prawer2008}.
Such a large scale integration opens the way to integrated all-optical photonic networks in diamond based on single colour centres implanted at well-defined positions within the PhC cavities.

\subsection*{Methods: Fabrication and patterning of the diamond membrane}
As starting material, we use heteroepitaxial diamond films with a thickness of about \unit[12]{$\mu$m} grown by chemical vapour deposition on Si(001) substrates via iri\-dium/yttria-stabilised zirconia (Ir/YSZ) buffer layers.
The Ir/YSZ/Si(001) substrates for diamond growth are prepared as described in \cite{Gsell2004}. A silicon wafer with a 4$^\circ$ off axis angle is used. To induce heteroepitaxial diamond nucleation the bias enhanced nucleation (BEN) procedure is applied. The nuclei with a typical areal density in the order of \unit[3$\cdot$10$^{11}$]{cm$^{-2}$} then evolve into oriented grains that merge during the formation of a closed layer. With increasing film thickness the mosaic spread drops down to few tenths of a degree. The grain coarsening continues until the network of small angle grain boundaries disintegrates and a dislocation-rich single crystal forms at a thickness above \unit[10]{$\mu$m} \cite{Schreck2001}. Plan view TEM images of our samples after PhC fabrication and electron diffraction patterns which show distinct diffraction maxima without any blurring confirm that this transition has taken place.

In order to fabricate a free-standing diamond membrane the silicon substrate is removed in  small areas (\unit[150]{$\mu$m} $\times$ \unit[150]{$\mu$m}) by a deep reactive ion etching (DRIE) process using an alternation of etching and passivation steps with SF$_6$ as an etching gas.
The Ir/YSZ buffer layers are removed from the nucleation side of the diamond film by ion beam etching with argon ions.
The diamond film  is thinned from  the nucleation side using reactive ion etching (RIE) with the following conditions:  15 sccm of oxygen gas, 10 sccm of argon gas, 5 sccm of SF$_6$
at a chamber pressure of 1 Pa.  Under these conditions the diamond film is thinned from \unit[12]{$\mu$m} to \unit[300]{nm} at a rate of \unit[114]{nm/min}. In this way the lower quality nucleation layer is completely removed and the cavity structure is placed in the highest quality layer at the surface.
The front side of the diamond film has been polished after the growth process to a rms-roughness of \unit[3]{nm}. The rms-roughness of the back side  is \unit[5]{nm} after the etching process.

In the case of the nanobeam cavities we deposit a \unit[80]{nm} thick titanium protection layer on top of the diamond membrane  prior to focused ion beam (FIB) milling in order to reduce unwanted surface milling due to beam tails.
The FIB milling is done in a Helios NanoLab 600 Dual beam (SEM/FIB) system  (FEI) with an energy of the Ga$^{+}$-ions of \unit[30]{keV} and a current of \unit[10]{pA}. An area of about \unit[10]{$\mu$m} $\times $ \unit[10]{$\mu$m} is exposed to the Ga$^{+}$-ions beam. We find that the FIB procedure partially damages the diamond material discernible from the occurrence of a graphite-like Raman signal at $\lambda \approx $ \unit[580]{nm} (excitation: $\lambda = $ \unit[532]{nm}).
Thus, after the focused ion beam milling, the sample is annealed at \unit[1000]{$^{\circ}$C} for 2 hours in vacuum (\unit[$10^{-3}$]{Pa}) to recover the pristine diamond film. Incorporated Ga-ions diffuse to the surface and out of the diamond film upon  high temperature annealing as confirmed by EDX measurements (see supplementary information).  Residual graphite layers are removed by conc. acid treatment (H$_2$SO$_4$:H$_2$O$_2$, 1:1 at room-temperature) and oxidation  in air (\unit[420]{$^{\circ}$C}, \unit[2]{h}). After the annealing and cleaning procedures no graphite-like Raman signal  can be detected on the patterned areas.


\clearpage
\subsection*{Acknowledgements}
We acknowledge helpful discussions with D. Englund.
We would like to thank J. Schmauch for SEM and TEM analysis, C. Zeitz for AFM measurements and F. Soldera for assistance with the FIB milling. Furthermore we cordially thank T. Jung for the fabrication tolerance analysis, K. Kretsch for assistance with the wet chemical etching, S. Griesing for sputtering of metal layers and S. Grandthyll for ellipsometry measurements. This work was financially supported by the Deutsche Forschungsgemeinschaft and the Bundesministerium f\"ur Bildung und Forschung (network EPHQAM, Contract No. 01Bl0903).
\subsection*{Authors Contribution}
J.R-M. and L.K. fabricated the photonic crystals, performed the experiments and did the numerical modeling of the structures.
M.F., S.G. and M.S. developed the CVD-growth process of the diamond films on iridium buffer layers. A.B. and M.W. prepared the diamond membrane.
J.R-M. and S.W. thinned the diamond film. C.P., J.R-M., L.K. and F.M. performed the FIB milling. C.H. and E.N. contributed experimental tools and helped with the PL-measurements and interpretation of data. C.B. conceived and designed the experiments.
J.R-M. and C.B. wrote the manuscript. All authors discussed the results and commented on the manuscript.
\subsection*{Competing Financial Interests}
The authors declare no competing financial interests.
\newpage
\newpage
\begin{figure}[h!]
  \centering
  \subfigure{
  \includegraphics[height=0.315\textheight]{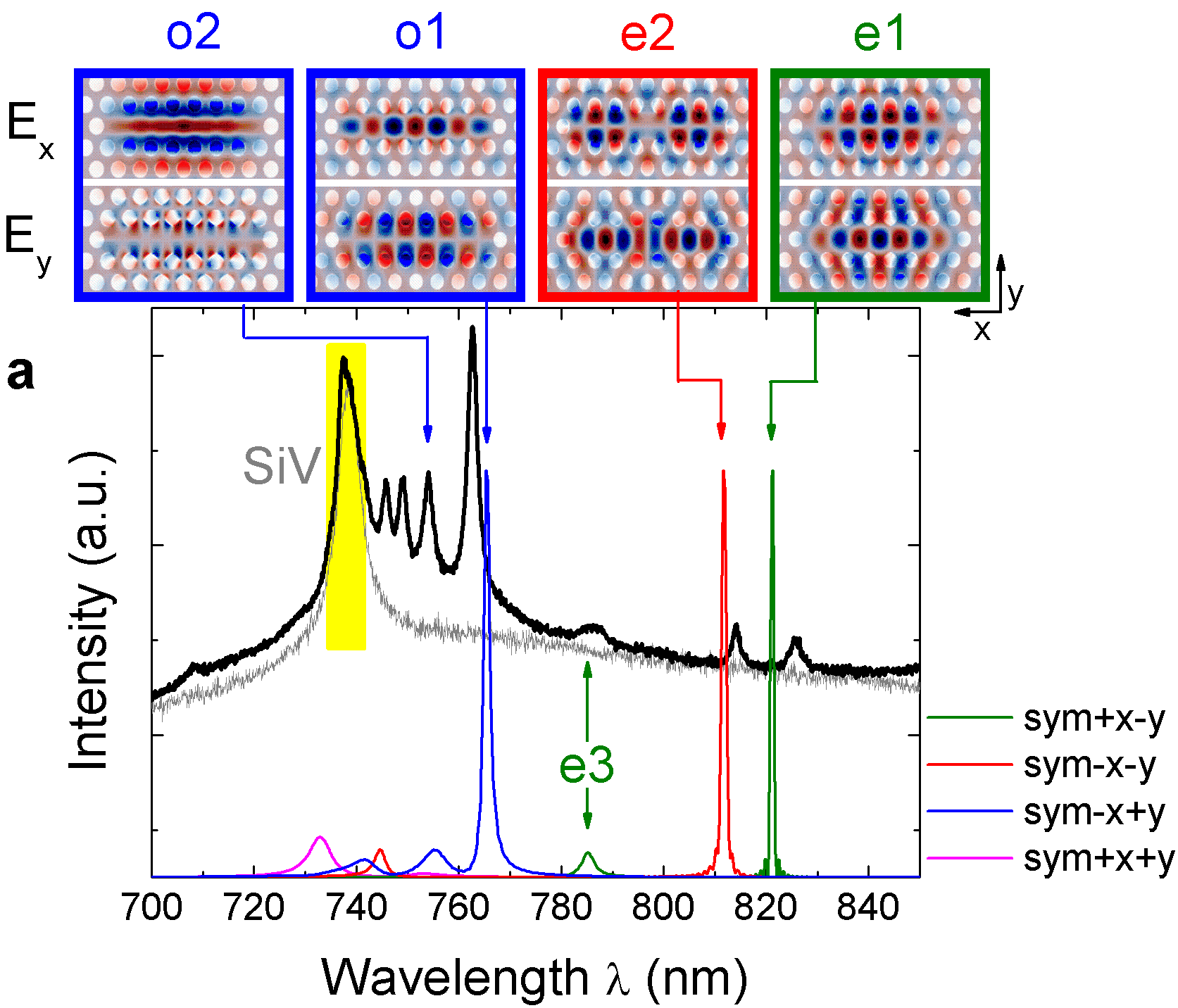}
  \label{fig:Modenbilder}}
  \subfigure{
  \includegraphics[height=0.315\textheight]{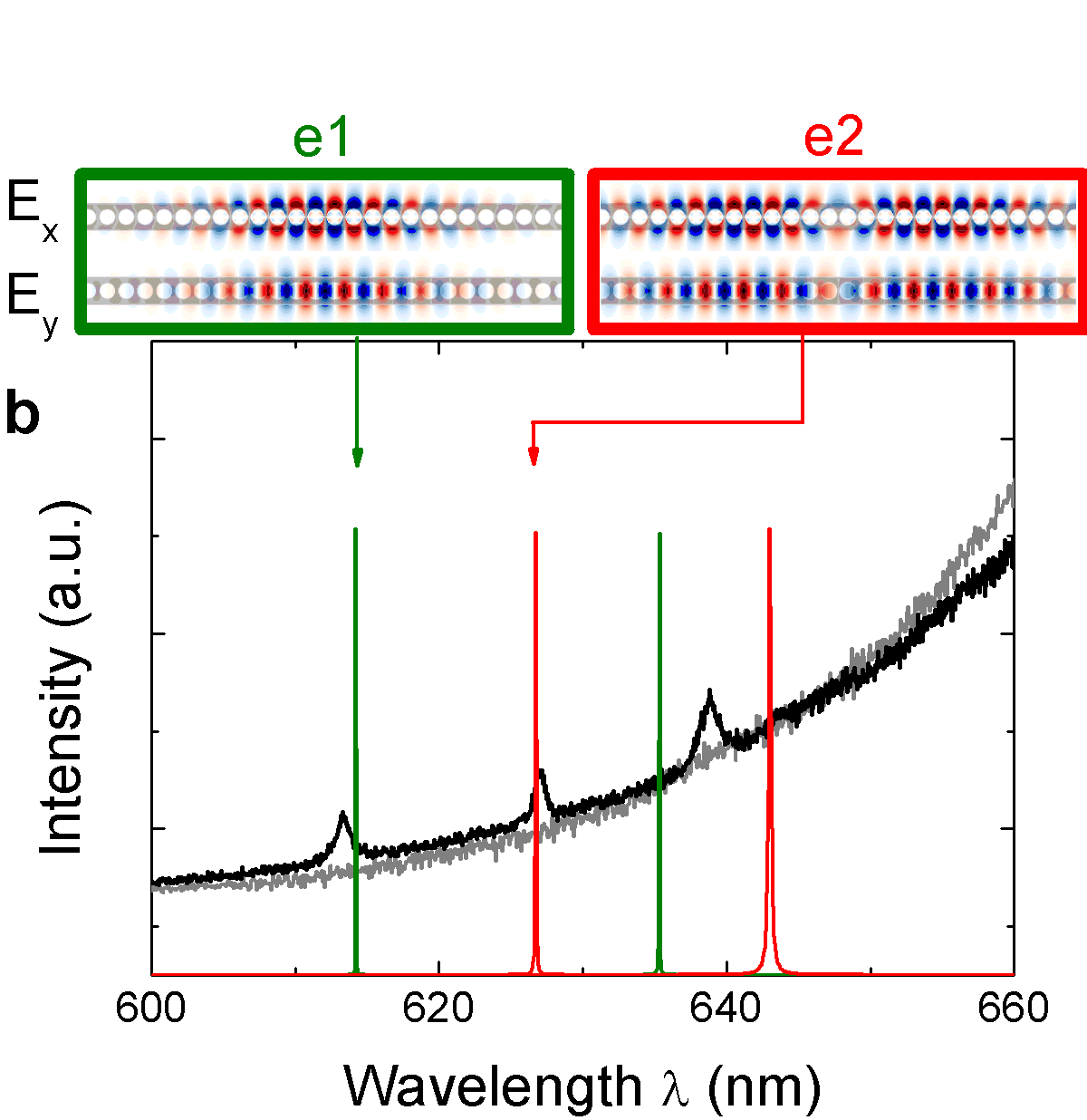}
  \label{fig:Nanobeam-spec}}
  \subfigure{
  \includegraphics[height=0.232\textheight]{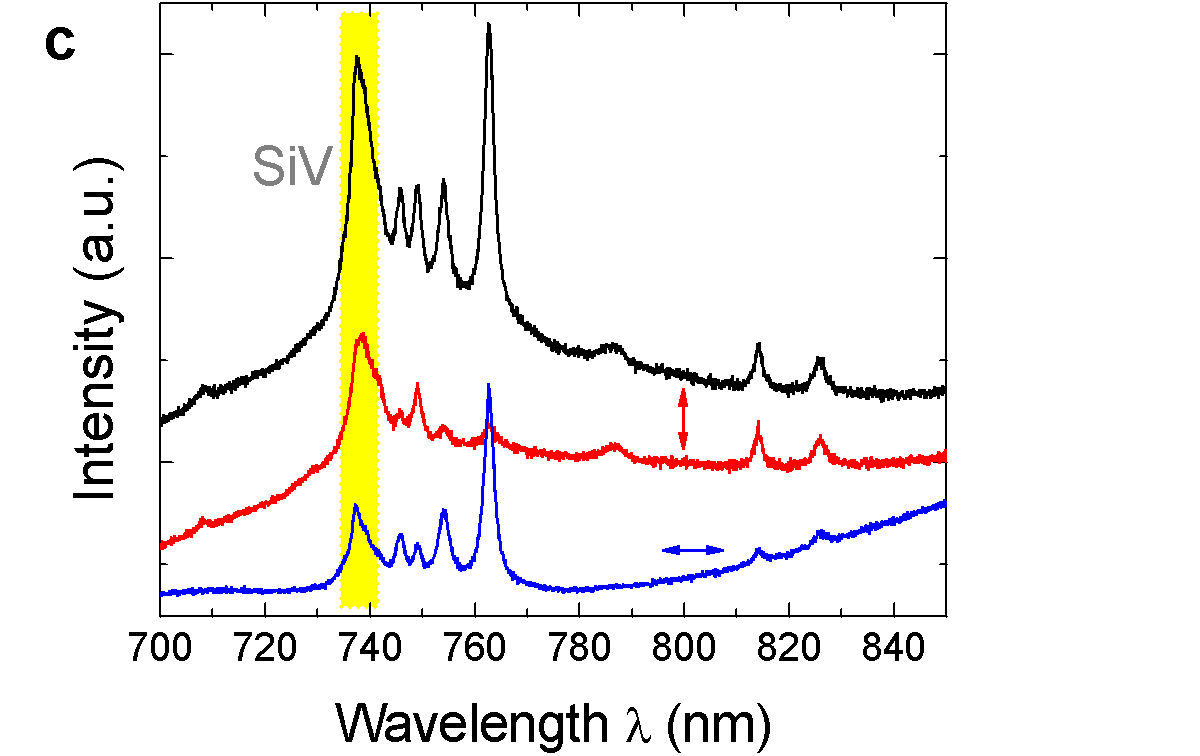}
  \label{fig:Polarisation}}
  \subfigure{
  \includegraphics[height=0.232\textheight]{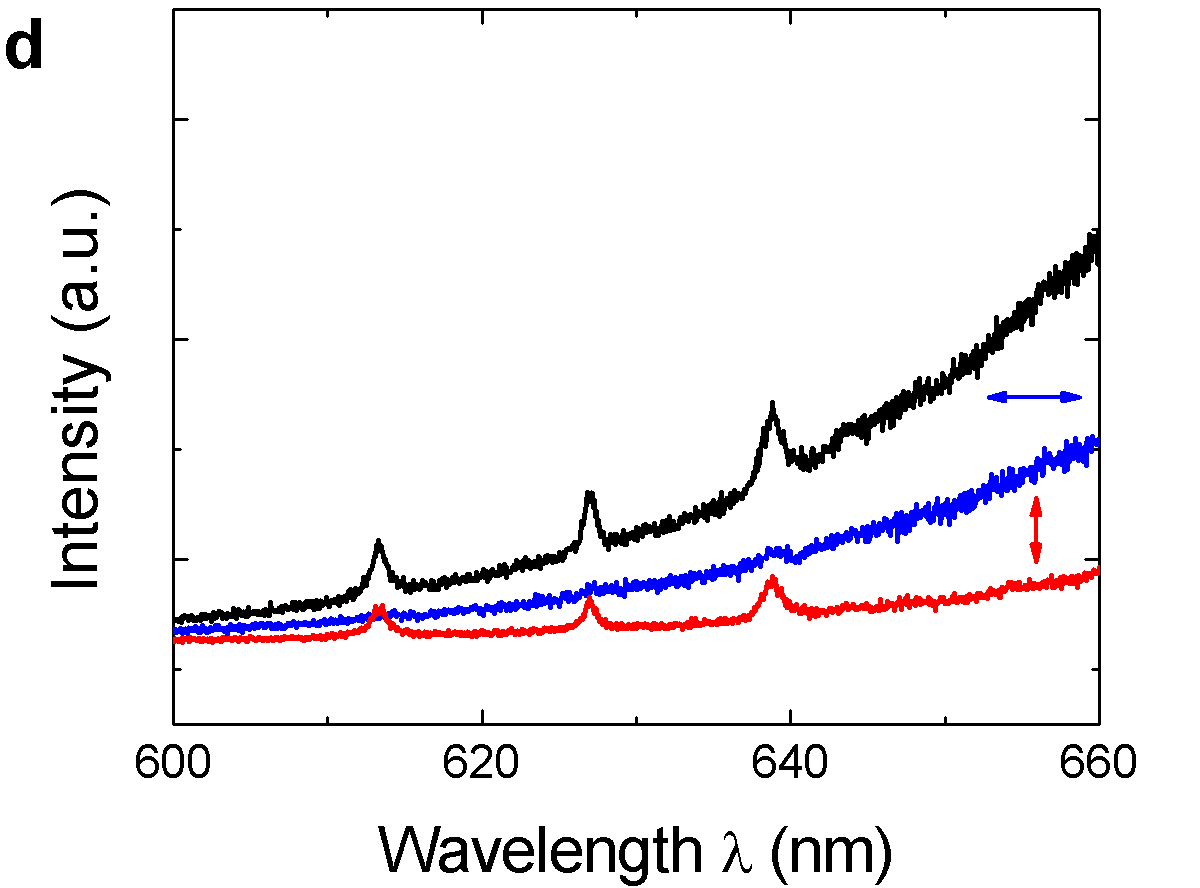}
  \label{fig:Nanobeam-pol}}
  \caption{\label{fig:Spektren}}
\end{figure}
\noindent
Figure \ref{fig:Spektren}: Experimental and simulated spectra: The experimental PL spectra of (a) a M7-cavity and (b) a nanobeam cavity  are shown in black and the reference spectra of the unstructured membrane are shown in gray. The  coloured curves show the simulated cavity modes where the different colours account for different symmetric boundary conditions applied in the simulation. The $E_x$ and $E_y$ mode profiles for the fundamental mode $e1$ and higher order modes $e2$, $o1$ and $o2$ are shown on top.
(a) The PL spectrum of the M7-cavity shows several cavity modes close to the SiV centre ZPL at $\lambda = $\unit[738]{nm} (yellow area). The intensity of the reference spectrum has to be scaled by a factor of six (see supplementary information).
 The simulated spectrum (arbitrary amplitude) of an  ideal M7-cavity with $R = 0.31a $
and $h = 1.1a$ matches the experimental results very well.
(b) In the PL spectrum of the nanobeam cavity three cavity modes are observed that are close to the design wavelength $\lambda = $ \unit[637]{nm} of the  NV$^{-}$ ZPL.  The calculated modes (arbitrary amplitude) of an ideal nanobeam cavity with  $h = w =1.5a$ whose radii decrease from $R = $ \unit[0.42]{$a$} at the cavity centre to $R = $ \unit[0.37]{$a$} at the structure edge, agree very well with the experimental measurement. Polarisation analysis of (c) a M7-cavity and (d) a nanobeam cavity: PL spectra with (without) polariser are shown by coloured (black) curves. The even modes are pronounced for a polariser oriented in the $y$-direction, whereas the odd modes are prominent for polariser oriented along the $x$-axis.
xis.

\begin{figure}[h!]
  \centering
  \subfigure{
  \includegraphics[height=0.2\textheight]{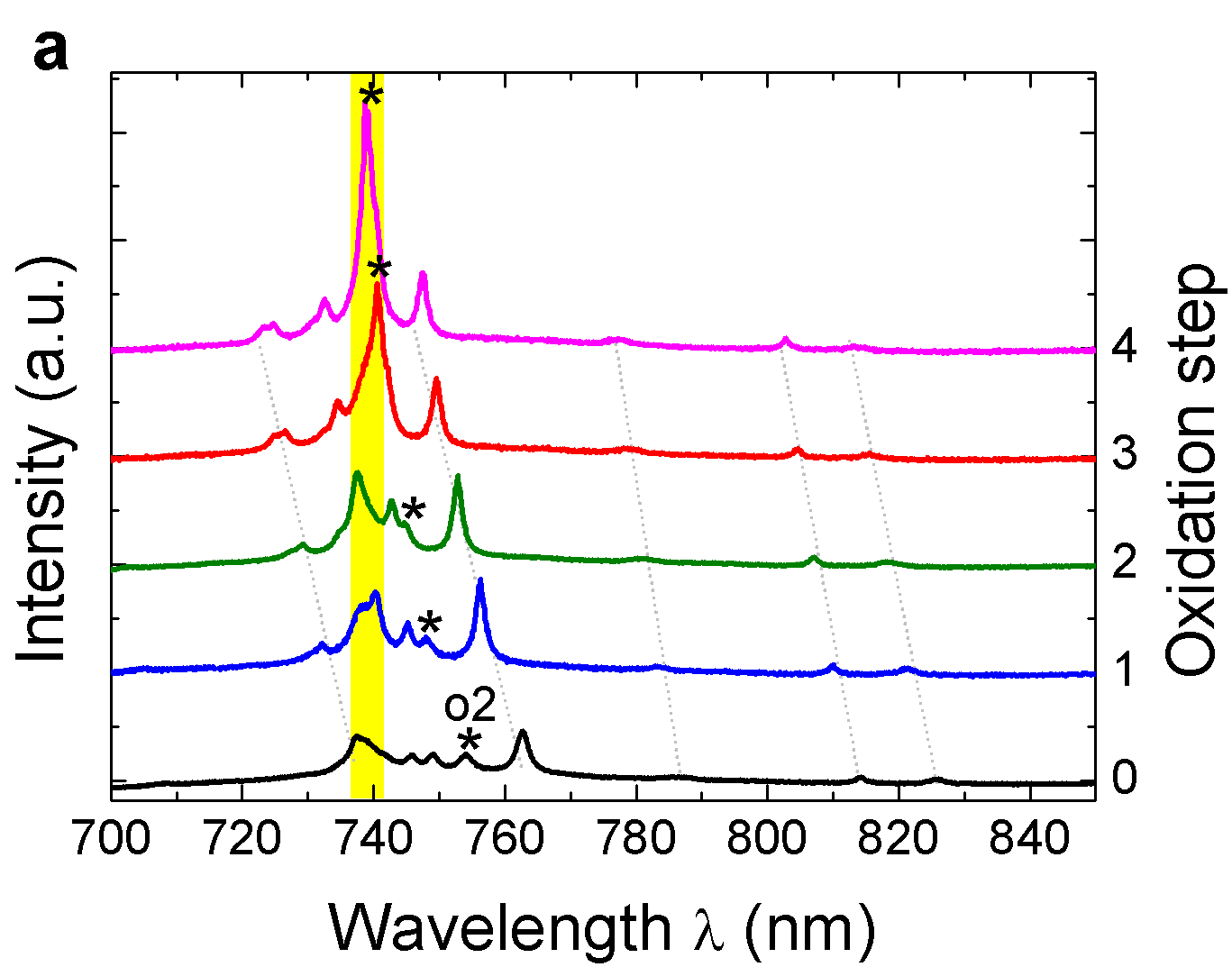}
  \label{fig:D115-CavityTuning}}
  \subfigure{
  \includegraphics[height=0.2\textheight]{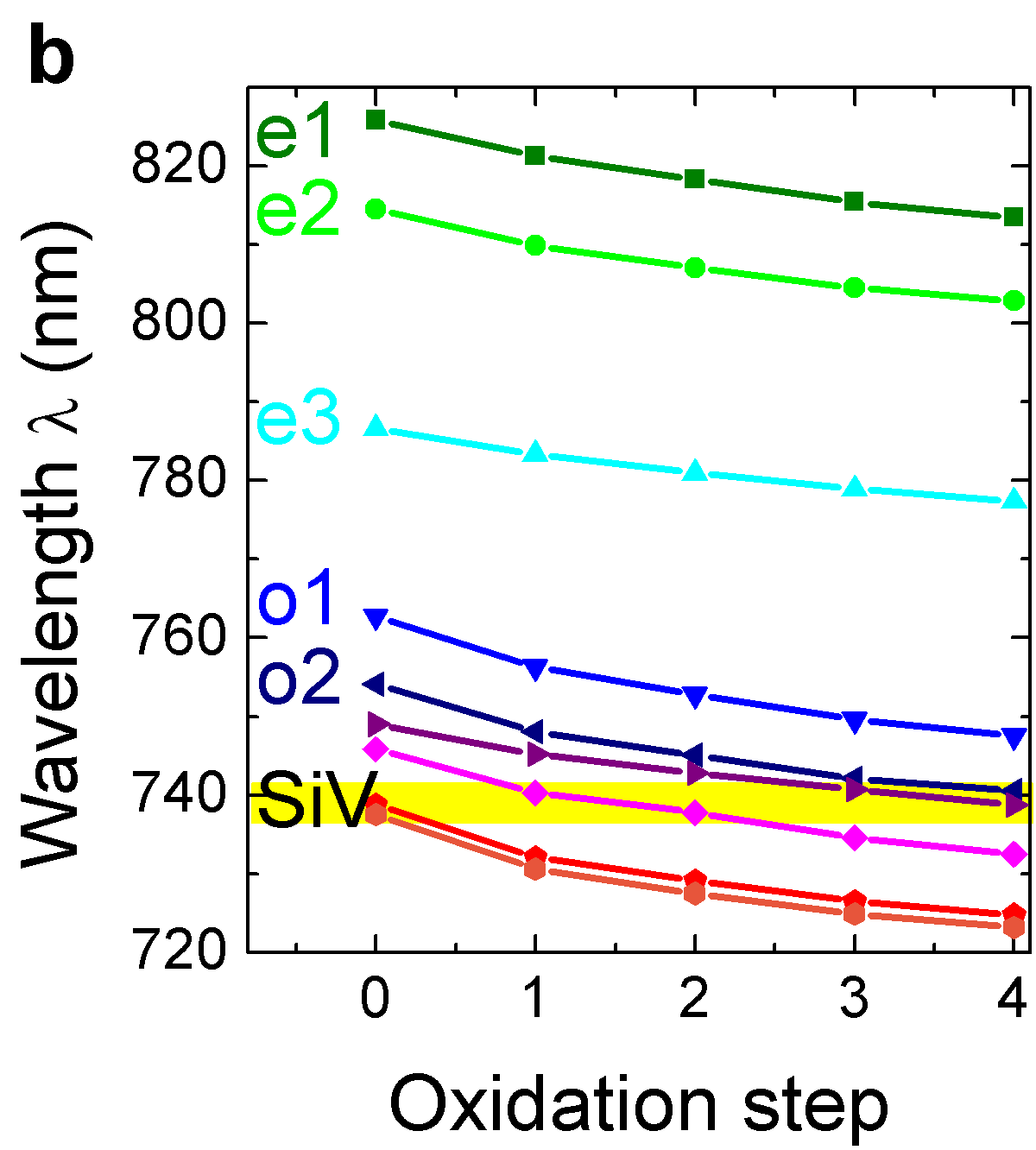}
  \label{fig:l-Oxid}}
  \subfigure{
  \includegraphics[height=0.2\textheight]{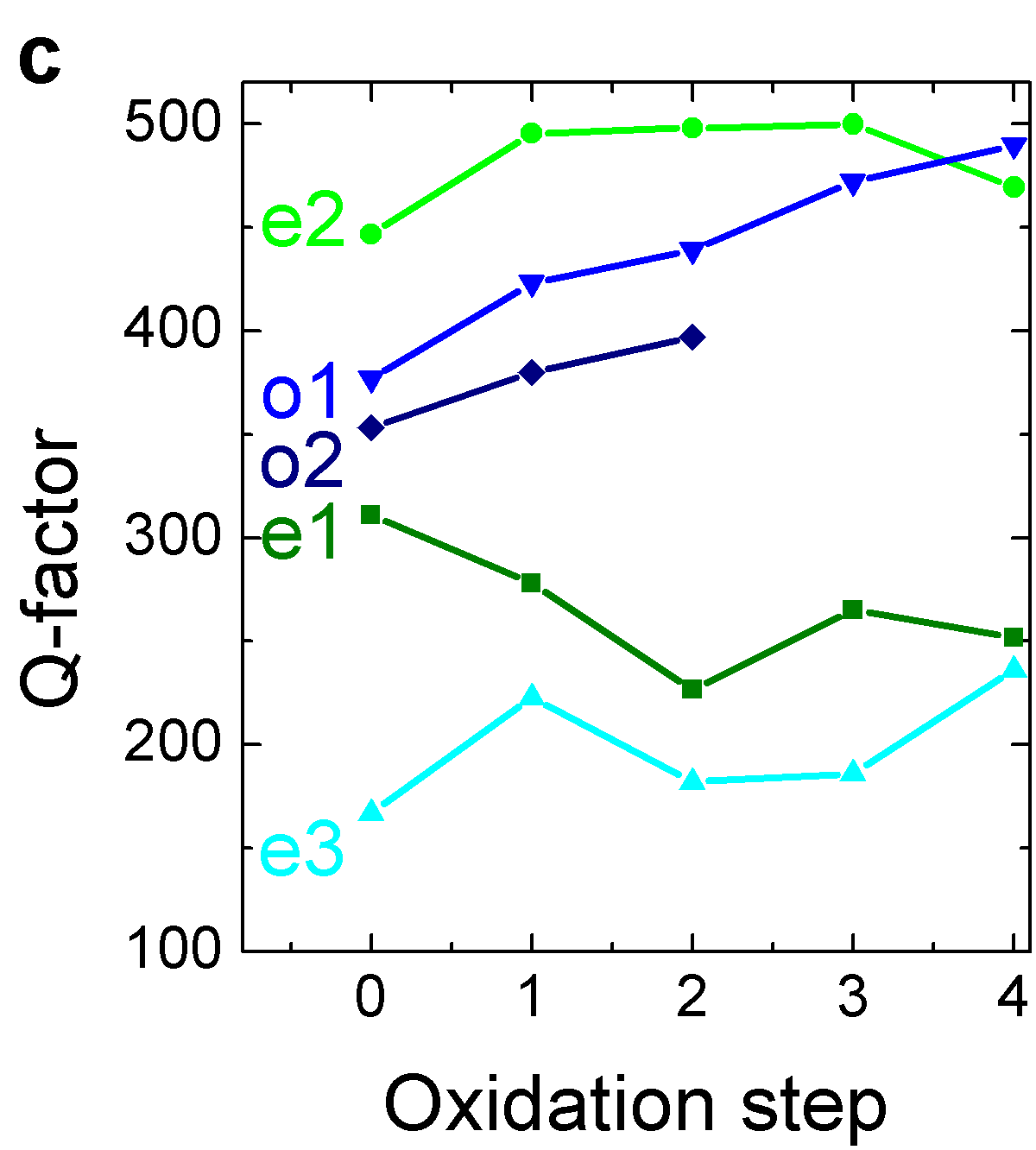}
  \label{fig:Q-Oxid}}
  \caption{Cavity tuning: (a) Cavity spectra taken after each oxidation step. When the cavity mode $o2$ (marked by *) is tuned into resonance with the emission line of  SiV centres, the intensity of the ZPL is clearly enhanced. (b) The resonant modes are blue-shifted by \unit[3]{nm} on average per oxidation step. In total the cavity modes are tuned up to  \unit[15]{nm}.  (c) The quality factors of the fundamental cavity modes show no significant degradation upon tuning.
  \label{fig:Cavity-tuning}}
\end{figure}

\begin{table}[h]
\centering
\begin{tabular}{|l||c|c|c|c|c|}
\hline
Modes & e1&  e2 & e3 & o1 & o2    \\
\hline \hline
Q$_{\text{Theory}}$ & 11,800 &  1,350 & 255 & 450 & 110  \\
\hline
Q$_{\text{SEM}}$    & 3,700  &  1,100 & 250 & 600 & 590  \\
\hline
Q$_{\text{Tilt}}$   & 900    &  770   & 250 & 600 & 115  \\
\hline
Q$_{\text{Exp}}$&  300 $\pm 50$&  450 $\pm 40$ & 180 $\pm 20$ & 380 $\pm 40$ & 400 $\pm 50$    \\
\hline
\end{tabular}
\caption{Theoretical and experimental Q-factors for the five lowest cavity modes: Here Q$_{\text{Theory}}$ denotes the calculated quality factor of an ideal M7-cavity, $Q_{\text{SEM}}$ the simulated Q-factor of the M7-cavity with a geometry based on the imported SEM image (assuming vertical sidewalls)  and Q$_{\text{Tilt}}$ the calculated Q-factor of a M7-cavity with ideal geometry but tilted sidewalls of 6$^{\circ}$. The hole taper primarily limits the quality factor Q$_{\text{Exp}}$ in the experiment. Q$_{\text{Exp}}$ values are averaged over 5 measurements.   \label{tab:Q-M7}}
\end{table}

\begin{table}[h]
\centering
\begin{tabular}{|l||c|c|c|}
\hline
Modes & e1&  e2 & e3    \\
\hline \hline
Q$_{\text{Theory}}$ & 540,000 &  260,000 & 23,000     \\
\hline
Q$_{\text{Exp}}$&       490 $\pm 25$&  700 $\pm 35$ & 470 $\pm 40$   \\
\hline
\end{tabular}
\caption{Theoretical and experimental Q-factors for the three lowest nanobeam modes: Here Q$_{\text{Theory}}$ denotes the quality factor of an ideal nanobeam-cavity and  Q$_{\text{Exp}}$ is the measured quality factor. Q$_{\text{Exp}}$ values are averaged over 5 measurements.
\label{tab:Q-Nanobeam}}
\end{table}
\end{document}